\documentclass[twocolumn,showpacs,unsortedaddress,superscriptaddress,showkeys,preprintnumbers,letterpaper]{revtex4-1}
\usepackage{soul}
\usepackage{acronym}
\usepackage{amsmath}
\usepackage{amssymb}
\usepackage{mathrsfs}
\usepackage{graphicx}
\usepackage{float}
\usepackage{appendix}
\usepackage{comment}
\usepackage{subfig}
\usepackage[usenames]{color}
\usepackage[normalem]{ulem}
\usepackage[bookmarksnumbered, bookmarksopen, breaklinks, colorlinks]{hyperref}
\pdfoutput=1

\newcommand{\AEI}{Albert-Einstein-Institut, Max-Planck-Institut f\"ur
Gravitationsphysik, D-30167 Hannover, Germany}
\newcommand{\Leibniz}{Leibniz Universit\"at Hannover, D-30167 Hannover, Germany}

\newcommand{\cross}{\times}
\newcommand{\RA}{\alpha}
\newcommand{\dec}{\delta}
\newcommand{\A}{\ensuremath{\mathcal{A}}}
\newcommand{\F}{\ensuremath{\mathcal{F}}}
\newcommand{\G}{\ensuremath{\mathcal{G}}}
\newcommand{\M}{\ensuremath{\mathcal{M}}}
\newcommand{\R}{\ensuremath{\mathcal{R}}}
\newcommand{\Y}{\ensuremath{Y}}
\newcommand{\mat}[1]{{\boldsymbol #1}}
\newcommand*{\ee}{\mathrm{e}}

\newcommand{\abs}[1]{\left\lvert #1 \right\rvert}

\begin{document}


\title{The multi-detector \F-statistic metric for short-duration non-precessing
inspiral gravitational-wave signals}

\author{Drew~Keppel}  
\email{drew.keppel@ligo.org}
\affiliation{\AEI}
\affiliation{\Leibniz}

\begin{abstract}
We derive explicit expressions for the multi-detector \F-statistic metric
applied to short-duration non-precessing inspiral signals. This is required for
template bank production associated with coherent searches for short-duration
non-precessing inspiral signals in gravitational-wave data from a network of
detectors. We compare the metric's performance with explicit overlap
calculations for all relevant dimensions of parameter space and find the metric
accurately predicts the loss of detection statistic above overlaps of 95\%. We
also show the effect that neglecting the variations of the detector response
functions has on the metric.
\end{abstract}

\maketitle
\acrodef{BNS}{binary neutron star}
\acrodef{CBC}{compact binary coalescence}
\acrodef{GW}{gravitational-wave}
\acrodef{PN}{post-Newtonian}
\acrodef{PSD}{power spectral density}
\acrodef{ROC}{Receiver Operator Characteristic}
\acrodef{SNR}{signal-to-noise ratio}
\acrodef{SPA}{stationary phase approximation}
\acrodef{SVD}{singular value decomposition}
\acrodef{IMR}{inspiral-merger-ringdown}
\acrodef{ASD}{amplitude spectral density}

\section{Introduction}

Inspiral signals are thought to be the most promising source of \acp{GW} for
second generation \ac{GW} detectors. Depending on the rate of merger events,
the Advanced LIGO, Advanced Virgo \ac{GW} detector network operating at design
sensitivity will be able to detect between 0.4 and 400 binary neutron star
coalescences per year~\cite{rates2010}. Underlying these numbers there is an
assumed threshold on the network \ac{SNR} at which a signal is ``detectable"
(i.e., has a false alarm probability below some established value).  It has
been shown that, among the different matched-filter based search strategies,
coherent templated searches for these signals can reduce the false alarm rate
for the same network \ac{SNR} compared to coincident templated
searches~\cite{Bose2000, Finn2001, Cutler2005, Harry2011}. Thus it is
attractive to prepare coherent searches for when the advanced detectors come
online in order to maximize the number of detected events.

The \F-statistic was originally derived as a single detector detection
statistic associated with searching \ac{GW} data for signals from rotating
neutron-stars~\cite{Jaranowski1998}, and was extended to multiple-detector
analysis in~\cite{Cutler2005}. However, it is equally applicable to coherent
searches for \ac{GW} signals from inspiralling compact objects~\cite{Bose2000,
Finn2001, Cutler2005, Harry2011}, due to the physical similarity of the two
emitting systems. The signals from both types of systems can be modelled as
\ac{GW} emission from a rotating quadrupole moment.  Both signals can be
characterized by four \emph{extrinsic} parameters that affect the amplitude,
polarization, and phase offset of the waveform, an extrinsic parameter that
sets a reference time for the signal, and \emph{intrinsic} parameters that
affect the phase and amplitude evolution of the waveform.

In performing templated matched-filter searches for \acp{GW}, one is always
faced with the question ``what template waveforms should the data be filtered
against?" With regards to searches for inspiral signals in single detector
\ac{GW} data, this question has been investigated within a geometric formalism.
Specifically, a distance measure can be defined on the parameter space based on
the ``mismatch" between waveforms from different parameter space
points~\cite{Owen:1995tm}. This was initially derived for the two dimensional
mass space for \ac{SPA} inspiral waveforms expanded to Newtonian order in the
amplitude and 1.0~\ac{PN} order in the phase, where the effects of the objects'
spins were neglected. This has been extended to 3.5~\ac{PN} order for the
``non-spinning" contributions to the phase~\cite{Owen:1998dk, Keppel2012}. In
addition, a higher dimensional metric has been obtained that includes the
``spin" contributions to the phase, up to 2.0~\ac{PN} order, for the case where
the objects' spins are aligned with the orbital angular
momentum~\cite{Brown2012}.

There have been several pieces of work that have been closely related to
deriving the multi-detector \F-statistic metric for short-duration
non-precessing inspiral signals.  The first was the derivation of the mismatch
metric for coherent searches of short-duration non-precessing inspiral signals
based purely on the Newtonian order inspiral phase model~\cite{Pai2000} and
built on the formalism of~\cite{Owen:1995tm, Bose2000}, which was later
extended to cover the phase expanded at 2.5~\ac{PN} order~\cite{Pai2001}.
Another was the derivation of the multi-detector \F-statistic metric for
rotating neutron-stars~\cite{PrixFstatMetric}.  In addition, there was the
computation of the Fisher matrix for the network \ac{SNR} of known and unknown
waveforms of short- and long-duration, focusing on obtaining explicit
expressions for the angular resolution of a \ac{GW} detector
network~\cite{wen2010}. Finally, the most closely related work showed parameter
recovery accuracies based on the Fisher matrix applied to inspiral and
inspiral-merger-ringdown waveforms observed by detector
networks~\cite{Ajith:2009fz}, although the derivation of the Fisher matrix was
was not presented. There has been no equivalent published derivation of the
multi-detector \F-statistic metric for short-duration non-precessing inspiral
signals including both the amplitude model, the phase model, and the
directional derivatives effects of detector responses. This is what we derive
here to 3.5~\ac{PN} order in the inspiral phase. This metric is required for
determining how to arrange templates that would cover the four dimensional
sky-location and mass space of a coherent search.

Previous coherent searches for short-duration non-precessing inspiral signals
have been based on one of three methods. They have either relied on the sky
position to be known precisely~\cite{GRBS6VSR23} or known to some degree and
tiled by detector triangulation arguments~\cite{Predoi2012}. These have both
done a templated search on the mass parameter space in an \emph{ad hoc} way
based on mass space coverings associated with a single
detector~\cite{Harry2011}. A third approach has been
hierarchical~\cite{Bose2011}, relying on coincident searches of single detector
data with their associated mass space coverings to decide what points in the
mass space are followed-up coherently. The metric derived here could be used as
the starting point for determining separately a template covering of the sky as
well as the mass space covering for a template bank associated with a coherent
search.

Following the formalism laid out for computing the multi-detector \F-statistic
in Refs.~\cite{PrixFstatMetric, PrixFstatTechNote}, this work is organized as
follows, Sec.~\ref{sec:signal} identifies the form of the \ac{GW} signal from
rotating non-precessing quadrupole moments as seen in a \ac{GW} network,
Sec.~\ref{sec:fstat} summarizes the formulation of the multi-detector
\F-statistic, Sec.~\ref{sec:approx} outlines the approximations appropriate
when applied to short-duration (i.e., much less than one day) non-precessing
inspiral signals, Sec.~\ref{sec:derivation} derives the metric for the coherent
multi-detector \F-statistic for short-duration non-precessing inspiral signals,
and Sec.~\ref{sec:verif} shows tests of this metric.

\section{Observed \ac{GW} signal from rotating non-precessing quadrupole moments}
\label{sec:signal}

To start with, let us identify the parameters that will affect how a generic
\ac{GW} signal from a rotating, non-precessing quadrupole moment is observed by
a \ac{GW} detector. These parameters can be separated into two classes,
\emph{intrinsic parameters}, which affect the time evolution of the waveform
and we will elaborate further on in Sec.~\ref{sec:approx}, and \emph{extrinsic
parameters}, which affect the polarization, amplitude, and the phase and time
offsets. The extrinsic parameters can be further subdivided into two classes,
those that can be measured analytically within the matched-filtering process,
and those that must be searched over by separate filters. As we will see, the
extrinsic parameters that can be measured analytically are the extrinsic
amplitude $h_0$, the inclination angle $\iota$ between the line of sight and
the total angular momentum of the emitting system, the reference phase
$\phi_0$, and the polarization angle $\psi$, which is a rotation between the
radiation frame of the \ac{GW} and the frame of the detector about the
direction of propagation $-\hat{n}$.

With those definitions of the extrinsic parameters, we give a signal model that
describes how a generic \ac{GW} signal from a rotating, non-precessing
quadrupole moment will be observed in detector~\Y. A generic propagating
\ac{GW} signal can be described in terms of two polarizations in general
relativity,
\begin{equation}
\mathrm{h} := h_{+} - i h_{\cross},
\end{equation}
where the $h_+$ and $h_\cross$ waveforms are out of phase by $90^\circ$. Thus,
these waveforms can be written in terms of the \emph{intrinsic waveforms}
$h_c(t)$ and $h_s(t)$ as
\begin{equation}
\mathrm{h} = \mathscr{A}_{+} h_c(t) - i \mathscr{A}_{\cross} h_s(t).
\end{equation}
The intrinsic waveforms can be further decomposed into an amplitude piece
$\mathsf{A}(t)$ and phase piece $\phi(t)$,
\begin{equation} \begin{aligned}
h_c(t) &:= \mathsf{A}(t) \cos[\phi(t) + \phi_0], \\
h_s(t) &:= \mathsf{A}(t) \sin[\phi(t) + \phi_0],
\end{aligned} \end{equation}
where $\mathsf{A}(t)$ and $\phi(t)$ will depend on the details of the emitting
system.  The \emph{polarization amplitudes} associated with the different
polarization waveforms are functions of the extrinsic amplitude $h_0$ and the
inclination angle $\iota$,
\begin{equation} \begin{aligned}
\mathscr{A}_+ := \frac{h_0}{2}(1 + \cos^2{\iota}), \;
\mathscr{A}_\cross := h_0 \cos{\iota}.
\end{aligned} \end{equation}

The waveform as seen by detector~\Y can be obtained by taking the real part of
this complex waveform projected onto the complex detector response
$\mathrm{F}^\Y := F_{+}^\Y + i F_{\cross}^\Y $,
\begin{equation}\label{eq:realdetectorsignal}
s^\Y(t) = \Re(\mathrm{h} \, \mathrm{F}^\Y e^{-i 2 \psi}),
\end{equation}
where we give explicit expressions for $F_{+}^\Y$ and $F_{\cross}^\Y$ later in
this section. Expanding $\mathrm{h}$, $\mathrm{F}$, and the phase terms of
cosine and sine waveforms of \eqref{eq:realdetectorsignal}, we find
\begin{multline}
s^\Y(t) = \\
\left(\mathscr{A}_+ \cos{\phi_0} \cos{2\psi} - \mathscr{A}_\cross \sin{\phi_0}
\sin{2\psi}\right) F_{+}^\Y(t) h_c^\Y(t) \\
+ \left(\mathscr{A}_+ \cos{\phi_0} \sin{2\psi} + \mathscr{A}_\cross
\sin{\phi_0} \cos{2\psi}\right) F_{\cross}^\Y(t) h_c^\Y(t) \\
+ \left(-\mathscr{A}_+ \sin{\phi_0} \cos{2\psi} - \mathscr{A}_\cross
\cos{\phi_0} \sin{2\psi}\right) F_{+}^\Y(t) h_s^\Y(t)\\
+ \left(-\mathscr{A}_+ \sin{\phi_0} \sin{2\psi} + \mathscr{A}_\cross
\cos{\phi_0} \cos{2\psi}\right) F_{\cross}^\Y(t) h_s^\Y(t).
\end{multline}
This can be separated into a sum over four detector-independent \emph{amplitude
parameters} $\{\A^\mu\}$ and four detector-dependent
\emph{polarization-weighted waveforms} $\{h_{\mu}^{\Y}(t)\}$,
\begin{equation}\label{eq:signal}
s^{\Y}(t) = \sum_{\mu = 1}^{4} \A^\mu h_{\mu}^{\Y}(t).
\end{equation}
It is readily apparent that the amplitude parameters are defined as
\begin{equation} \begin{aligned}
\A^1 &:= \;\;\; \mathscr{A}_+ \cos{\phi_0} \cos{2\psi} - \mathscr{A}_\cross
\sin{\phi_0} \sin{2\psi}, \\
\A^2 &:= \;\;\; \mathscr{A}_+ \cos{\phi_0} \sin{2\psi} + \mathscr{A}_\cross
\sin{\phi_0} \cos{2\psi}, \\
\A^3 &:= -\mathscr{A}_+ \sin{\phi_0} \cos{2\psi} - \mathscr{A}_\cross
\cos{\phi_0} \sin{2\psi}, \\
\A^4 &:= -\mathscr{A}_+ \sin{\phi_0} \sin{2\psi} + \mathscr{A}_\cross
\cos{\phi_0} \cos{2\psi},
\end{aligned} \end{equation}
while the polarization-weighted waveforms are defined as
\begin{equation} \begin{aligned}
h_1^\Y(t) &:= F_+^{\Y}(t) h_c(t - t^\Y), \\
h_2^\Y(t) &:= F_\cross^{\Y}(t) h_c(t - t^\Y), \\
h_3^\Y(t) &:= F_+^{\Y}(t) h_s(t - t^\Y), \\
h_4^\Y(t) &:= F_\cross^{\Y}(t) h_s(t - t^\Y).
\end{aligned} \end{equation}

Turning our attention to the detector polarization responses, these
characterize the response of an arbitrary \ac{GW} detector for signals that
satisfy the long wavelength limit approximation~\cite{Rakhmanov2008}. They can
be defined as the double contraction of two tensors~\cite{PrixFstatTechNote},
\begin{equation}\label{eq:FpFx}
F_+^{\Y}(t) := \epsilon_+^{i j} d^\Y_{i j}(t), \;
F_\cross^{\Y}(t) := \epsilon_\cross^{i j} d^\Y_{i j}(t), \end{equation}
where $d^\Y_{i j}(t)$ is the \emph{detector response tensor} and
$\{\epsilon_{+,\cross}^{i j}\}$ are the \emph{polarization-independent basis
tensors of the radiation frame}. For an interferometric detector, the detector
response tensor is given by
\begin{equation}
d^\Y_{i j}(t) = \frac{1}{2} \left\{ \hat{l}^\Y_1(t) \otimes \hat{l}^\Y_1(t) -
\hat{l}^\Y_2(t) \otimes \hat{l}^\Y_2(t) \right\}_{i j}.
\end{equation}
Here, $\hat{l}^\Y_1$ is the unit vector pointing along interferometer~\Y's
first arm away from the interferometer's vertex. Similarly, $\hat{l}^\Y_2$ is
the unit vector pointing along interferometer~\Y's second arm away from the
interferometer's vertex. The polarization-independent basis tensors are defined
as
\begin{equation} \begin{aligned}
\epsilon^{i j}_+ &:= \left\{\hat{\xi} \otimes \hat{\xi} - \hat{\eta} \otimes
\hat{\eta}\right\}^{i j}, \\
\epsilon^{i j}_\cross &:= \left\{\hat{\xi} \otimes \hat{\eta} + \hat{\eta}
\otimes \hat{\xi}\right\}^{i j},
\end{aligned} \end{equation}
given in the radiation frame $\{\hat{\xi}, \hat{\eta}, -\hat{n}\}$, where
$-\hat{n}$ is the direction of propagation, and $\{\hat{\xi}, \hat{\eta}\}$ are
basis vectors in the wave-plane (i.e., the plane perpendicular to direction of
propagation).  The basis vectors $\hat{\xi}$ and $\hat{\eta}$ can be defined
with respect to $\hat{n}$ as
\begin{equation} \begin{aligned}
\hat{\xi} := \frac{\hat{n} \cross \hat{z}}{\abs{\hat{n} \cross \hat{z}}}, \;
\hat{\eta} := \hat{\xi} \cross \hat{n}.
\end{aligned} \end{equation}
In a fixed reference frame centered at the geocenter, where
\begin{equation}
\hat{n} = (\cos{\dec} \cos{\RA}, \cos{\dec} \sin{\RA}, \sin{\dec}),
\end{equation}
the wave-plane basis vectors are
\begin{gather}
\hat{\xi} = (\sin{\RA}, -\cos{\RA}, 0), \\
\hat{\eta} = (-\sin{\dec} \cos{\RA}, -\sin{\dec} \sin{\RA}, \cos{\dec}).
\end{gather}

We have now defined all of the quantities that are used to convert a \ac{GW}
signal from an arbitrary non-precessing rotating quadrupole moment source to the
signal seen by a \ac{GW} detector. The remaining details of the signal will
depend on the specifics of the emitting system.

\section{The \F-statistic}
\label{sec:fstat}

The likelihood ratio of a signal $\mat{s}$ being in the data of a network of
\ac{GW} detectors $\mat{x}$ is given as
\begin{equation}\label{eq:likelihood}
\Lambda(\mat{x}; \mat{s}) = \frac{P(\mat{x}|\mat{s})}{P(\mat{x}|0)} =
\exp\left[ (\mat{x}|\mat{s}) - \frac{1}{2} (\mat{s}|\mat{s})\right],
\end{equation}
where $(\mat{a}|\mat{b}) := \sum_{\Y} (a^\Y|b^\Y)$ and the definition of the
noise-weighted inner product $(a^\Y|b^\Y)$ depends on the details of the
waveform being studied. We define this for inspiral signals in~\eqref{eq:inner}
of Sec.~\ref{sec:approx}. Using the signal model from \eqref{eq:signal},
\eqref{eq:likelihood} can be written as
\begin{equation}\label{eq:lnlikelihood}
\ln \Lambda(\mat{x}; \mat{s}) = \A^{\mu} x_\mu - \frac{1}{2} \A^\mu \M_{\mu
\nu} \A^{\nu},
\end{equation}
where $x_\mu := (\mat{x}|\mat{h}_\mu)$ and $\M_{\mu \nu} :=
(\mat{h}_\mu|\mat{h}_\nu)$. In matrix form, $\M_{\mu \nu}$ is block diagonal,
\begin{equation}
\M_{\mu \nu} = \left(\begin{array}{cccc}
A & C & 0 & 0 \\
C & B & 0 & 0 \\
0 & 0 & A & C \\
0 & 0 & C & B \\
\end{array}\right)_{\mu \nu},
\end{equation}
due to the orthogonality of the sine and cosine intrinsic waveforms. Here,
\begin{subequations}\begin{equation}
A := \sum_\Y F_+^\Y(t) F_+^\Y(t) (h_c^\Y|h_c^\Y),
\end{equation} \begin{equation}
B := \sum_\Y F_\cross^\Y(t) F_\cross^\Y(t) (h_c^\Y|h_c^\Y),
\end{equation} \begin{equation}
C := \sum_\Y F_+^\Y(t) F_\cross^\Y(t) (h_c^\Y|h_c^\Y).
\end{equation} \end{subequations}

The log likelihood ratio of \eqref{eq:lnlikelihood} can be analytically maximized
over the amplitude parameters, resulting in the maximum likelihood ratio
\emph{\F-statistic}~\cite{Cutler2005},
\begin{equation}\label{eq:fstat}
\F := \ln \Lambda(\mat{x}; \mat{s}_{\rm ML}) = \frac{1}{2} x_\mu \M^{\mu \nu} x_\nu,
\end{equation}
where $\M^{\mu \nu} := \{\M^{-1}\}^{\mu \nu}$ is the inverse of $\M_{\mu \nu}$,
i.e.~$\M^{\mu \alpha} \M_{\alpha \nu} = \delta^\mu_\nu$, and takes the
following form,
\begin{equation}\label{eq:Minverse}
\M^{\mu \nu} = \frac{1}{D}\left(\begin{array}{cccc}
B & -C & 0 & 0 \\
-C & A & 0 & 0 \\
0 & 0 & B & -C \\
0 & 0 & -C & A \\
\end{array}\right)^{\mu \nu},
\end{equation}
where $D := AB - C^2$.  It should be noted that the \F-statistic is the same as
the square of coherent \ac{SNR} ($2\F = \rho^2_{\textrm{coh}}$), which has been
previously used in literature associated with coherent searches for inspiral
signals with ground-based \ac{GW} detectors~\cite{Cutler2005, Harry2011}.

\section{Application to Inspiral Signals}
\label{sec:approx}

So far our treatment of the \F-statistic could be equally applied to searching
for \ac{GW} signals from rotating neutron stars or inspiralling binaries of
compact objects.  Restricting ourselves to the case of non-spinning inspiral
signals, the intrinsic parameters include $\{\eta, \M_c\}$, where $\eta := m_1
m_2 / (m_1 + m_2)^2$ is the symmetric mass ratio and $\M_c := (m_1 + m_2)
\eta^{3/5}$ is the chirp mass. In addition, there is an extrinsic parameter
that can be efficiently maximized over but has not been in deriving the
\F-statistic, namely the coalescence time $t_c$.  This can be easily done with
the use of the Fast Fourier Transform.  The additional extrinsic parameters
that must be searched over with separate filters are the sky locations $\{\RA,
\dec\}$, where $\RA$ is the right ascension and $\dec$ is the declination.

Although we restrict the derivation to the case of short-duration non-spinning
inspiral signals, spins aligned with the angular momentum could easily be
incorporated into the phase model and included as intrinsic parameters. This is
because binaries in which the objects' spins are aligned with the angular
momentum do not precess.

In second generation \ac{GW} detectors, the sensitive band of the detectors
will start as low as 10Hz. A binary neutron star system's \ac{GW} signal will
be in the sensitive band of the detectors for $\lesssim 17$ minutes before
coalescing, which amounts to a rotation of the Earth of $\lesssim
0.07$~radians. Thus, for a source's fixed sky location, the detectors can
approximated as fixed $d^\Y_{i j}(t) \approx d^\Y_{i j}(t_c)$. With this
approximation, the polarization weighted waveforms are given in the frequency
domain as
\begin{equation}\begin{aligned}
h_1^\Y(f) &:= F_+^{\Y} h_c(f), \\
h_2^\Y(f) &:= F_\cross^{\Y} h_c(f), \\
h_3^\Y(f) &:= F_+^{\Y} h_s(f), \\
h_4^\Y(f) &:= F_\cross^{\Y} h_s(f).
\end{aligned}\end{equation}
The frequency domain intrinsic waveforms are given as
\begin{equation}\begin{aligned}
h_c(f) &:= \mathrm{A}(f) \Re \ee^{i\Psi(f)}, \\
h_s(f) &:= \mathrm{A}(f) \Im \ee^{i\Psi(f)},
\end{aligned}\end{equation}
where $\mathrm{A}(f)$ is the intrinsic amplitude of the waveform, $\Psi(f)$ is
the phase of the waveform, and $\Re$ and $\Im$ denote operators that extract
the real and imaginary parts, respectively. Each of these components has the
following dependencies
\begin{equation}\begin{gathered}\label{eq:intrinsicparamdeps}
F_+^{\Y} = F_+^{\Y}(t_c; \RA, \dec), \\
F_\cross^{\Y} = F_\cross^{\Y}(t_c; \RA, \dec), \\
\mathrm{A}(f) = \mathrm{A}(f; \M_c, \eta), \\
\Psi(f) = \Psi(f; t_c, \RA, \dec, \M_c, \eta).
\end{gathered}\end{equation}
NB: $\mathrm{A}(f)$ is typically defined to include $h_0$ and $\Psi(f)$ is
typically defined (e.g., \cite{Keppel2012}) to include $\phi_0$, however in
this treatment, $h_0$ and $\phi_0$ are instead included as part of the
amplitude parameters. The explicit expressions of $\mathrm{A}(f)$ and $\Psi(f)$
according to the Stationary Phase Approximation are expanded to Newtonian order
in the amplitude and 3.5~\ac{PN} order in the phase in
Appendix~\ref{app:PNwaveform}.

The template waveforms for inspiral signals occupy a large bandwidth within the
detectors, entering the sensitive band at the lower frequency cutoff $f_{\rm
low}$ and extending up to the frequency associated with the inner-most stable
circular orbit $f_{\rm ISCO}$. For these signals, the inner product between two
waveforms is defined as
\begin{equation}\label{eq:inner}
(x^\Y|y^\Y) := 4 \Re \int\limits_{f_{\rm low}}^{f_{\rm high}}
\frac{\tilde{x}^\Y(f) \tilde{y}^{\Y*}(f)}{S^{\Y}(f)} df,
\end{equation}
where $f_{\rm high}$ is the upper cutoff frequency given by the smaller of the
Nyquist frequency of that data or $f_{\rm ISCO}$, $\tilde{x}(f)$ denotes the
Fourier transform of $x(t)$, $(.)^*$ denotes the complex conjugate operator,
and $S^{\Y}(f)$ is the one-sided \ac{PSD} of detector~\Y.

\section{\F-statistic metric derivation}
\label{sec:derivation}

The metric on the full set of parameters $\{\lambda\}$ including intrinsic and
extrinsic parameters can be derived by expanding the log likelihood ratio
\eqref{eq:likelihood} for a mismatched signal to second order in the parameter
differences, $\Delta \lambda$,
\begin{multline}\label{eq:expandedlikelihood}
2 \ln \Lambda(\mat{s(\lambda)};\mat{s}(\lambda + \Delta \lambda)) = \\
(\mat{s(\lambda)}|\mat{s}(\lambda)) -
(\partial_a\mat{s(\lambda)}|\partial_b\mat{s}(\lambda)) \Delta \lambda^a \Delta
\lambda^b \\
 + \mathcal{O}(\Delta \lambda^3),
\end{multline}
where $\partial_a := \partial / \partial \lambda^a$ is the partial derivative
w.r.t.~parameter $\lambda^a$. In this notation, we will restrict the use of
greek indices to the amplitude parameters (i.e., $\lambda^\mu = \A^{\mu}$) and
for the metric subspace associated with the amplitude parameters. Using
\eqref{eq:expandedlikelihood}, we are led to the definition of the full metric
$g_{a b}$, which measures the fractional loss of 2\F, as
\begin{equation}
g_{a b} := \frac{(\partial_a \mat{s}|\partial_b \mat{s})}{(\mat{s}|\mat{s})}.
\end{equation}
Recalling that $(\partial_\mu \A^\alpha \mat{h}_\alpha| \partial_\nu \A^\beta
\mat{h}_\beta) = (\mat{h}_\mu|\mat{h}_\nu) = \M_{\mu \nu}$, and using the
signal model from \eqref{eq:signal}, this metric can be decomposed into blocks
\begin{equation}\label{eq:fullmetric}
g_{a b} = \frac{1}{\A^\alpha \M_{\alpha \beta} \A^\beta}\left(
\begin{array}{cc} \M_{\mu \nu} & \A^\alpha \R_{\mu \alpha j} \\ \A^\beta
\R_{\nu \beta i} & \A^\alpha h_{\alpha \beta i j} \A^\beta \end{array}
\right)_{ab}.
\end{equation}
As stated before, in the above equation, the indices $\mu$ and $\nu$ are
associated with the amplitude parameter subspace and the indices $i$ and $j$
are associated with the non-amplitude parameter subspace. The quantities
$\R_{\mu \nu i}$ and $h_{\mu \nu i j}$ are defined as 
\begin{align}
\R_{\mu \nu i} &:= (\mat{h}_\mu|\partial_i \mat{h}_\nu), \\
h_{\mu \nu i j} &:= (\partial_i \mat{h}_\mu|\partial_j \mat{h}_\nu)
\end{align}
The $\M_{\mu\nu}$ block is associated with derivatives of only the amplitude
parameter subspace, the $\A^\alpha h_{\alpha \beta i j} \A^\beta$ block only
with derivatives of the non-amplitude parameter subspace, and the $\A^\alpha
\R_{\mu \alpha i}$ block with derivatives of both subspaces.

To obtain the metric for the \F-statistic, we can project out the dimensions
associated with the amplitude subspace~\cite{Krolak2004, PrixFstatMetric}
\begin{equation}
g^{\F}_{i j} = g_{i j} - g_{i \alpha} g^{\alpha \beta} g_{\beta j}.
\end{equation}
Using the form of the full metric from \eqref{eq:fullmetric}, the \F-statistic
metric can be written as
\begin{equation}\label{eq:fstatmetric}
g^{\F}_{i j} = \frac{\A^\alpha \G_{\alpha \beta i j} \A^\beta}{\A^\alpha
\M_{\alpha \beta} \A^\beta},
\end{equation}
where the projected Fisher matrix $\G_{\mu \nu i j}$ is given by
\begin{equation}
\G_{\mu \nu i j} = h_{\mu \nu i j} - \R_{\alpha \mu i} \M^{\alpha \beta}
\R_{\beta \nu j}.
\end{equation}
The two pieces of the projected Fisher matrix we refer to as the non-amplitude
parameter subspace matrix, $h_{\mu \nu i j}$, and the amplitude subspace
maximization correction, $\R_{\alpha \mu i} \M^{\alpha \beta} \R_{\beta \nu
j}$.  Similar to the derivation in Appendix B of Ref.~\cite{PrixFstatMetric},
after symmetrizing on $(\mu,\nu)$ and $(i,j)$, $h_{\mu \nu i j}$ takes the form
\begin{equation}\label{eq:P}
h_{\mu \nu i j} = \left(\begin{array}{cccc}
P^1_{ij} & P^3_{ij} & 0 & P^4_{ij} \\
P^3_{ij} & P^2_{ij} & -P^4_{ij} & 0 \\
0 & -P^4_{ij} & P^1_{ij} & P^3_{ij} \\
P^4_{ij} & 0 & P^3_{ij} & P^2_{ij} \end{array}\right)_{\mu \nu}.
\end{equation}
Although this looks identical to the derivation for rotating neutron star
signals~\cite{PrixFstatMetric}, one difference to keep in mind is that for
inspiral signals there are additional terms hidden in these components. These
additional terms are the result of the presence of an intrinsic parameter in
the amplitude of the signal, what we refer to as the \emph{intrinsic
amplitude}.  This can be seen in \eqref{eq:intrinsicparamdeps}.  The components
of $h_{\mu \nu i j}$ are given as
\begin{subequations}\begin{equation}
P^1_{ij} = \mat{f}^{++} \cdot \mat{G}_{ij}
+ \mat{f}^{++}_{i} \cdot \mat{J}_j + \mat{f}^{++}_{j} \cdot \mat{J}_i
+ \mat{f}^{++}_{ij} \cdot \mat{H},
\end{equation}\begin{equation}
P^2_{ij} = \mat{f}^{\cross\cross} \cdot \mat{G}_{ij}
+ \mat{f}^{\cross\cross}_{i} \cdot \mat{J}_j + \mat{f}^{\cross\cross}_{j} \cdot
\mat{J}_i
+ \mat{f}^{\cross\cross}_{ij} \cdot \mat{H},
\end{equation}\begin{multline}
P^3_{ij} = \mat{f}^{+\cross} \cdot \mat{G}_{ij} \\
+ \frac{1}{2}\left(\mat{f}^{+\cross}_i + \mat{f}^{\cross+}_i\right) \cdot
\mat{J}_j + \frac{1}{2}\left(\mat{f}^{+\cross}_j + \mat{f}^{\cross+}_j\right)
\cdot \mat{J}_i \\
+ \frac{1}{2}\left(\mat{f}^{+\cross}_{ij} + \mat{f}^{\cross+}_{ij}\right) \cdot
\mat{H},
\end{multline}\begin{equation}
P^4_{ij} = \frac{1}{2}\left(\mat{f}^{+\cross}_i - \mat{f}^{\cross+}_i\right)
\cdot \mat{K}_j + \frac{1}{2}\left(\mat{f}^{+\cross}_j -
\mat{f}^{\cross+}_j\right) \cdot \mat{K}_i.
\end{equation}\end{subequations}
Above, we have introduced the detectors' polarization response vectors
$\mat{f}_{(ij)}^{pq}$, the detectors' waveform vectors $\{\mat{H}, \mat{J}_{i},
\mat{K}_{i}, \mat{G}_{ij}\}$, and the notation $\mat{x} \cdot \mat{y} :=
\sum_{\Y} x^\Y y^\Y$, which denotes a sum over detectors,. The detectors'
polarization response vectors are defined as
\begin{equation}
f^{pq\Y} := F_{p}^{\Y}F_{q}^{\Y},
\end{equation}\begin{equation}
f^{pq\Y}_i := \partial_i F_{p}^{\Y} F_{q}^{\Y},
\end{equation}\begin{equation}
f^{pq\Y}_{ij} := \partial_i F_{p}^{\Y} \partial_j F_{q}^{\Y},
\end{equation}
where the derivatives of the detector polarization responses $\partial_i
F_{+,\cross}^\Y$ are given in Appendix~\ref{app:detpolrespderivs}. Next, the
detectors' waveform vectors $\{\mat{H}, \mat{J}_{i}, \mat{K}_{i},
\mat{G}_{ij}\}$ are defined as
\begin{equation}
G_{ij}^\Y := (h^\Y \partial_i \ln \mathrm{A}|h^\Y \partial_j \ln \mathrm{A}) +
(h^\Y \partial_i \Psi^\Y(f)|h^\Y \partial_j \Psi^\Y(f)),
\end{equation}\begin{equation}
H^\Y := (h^\Y |h^\Y),
\end{equation}\begin{equation}
J_{i}^\Y := (h^\Y |h^\Y \partial_i \ln \mathrm{A}),
\end{equation}\begin{equation}
K_{i}^\Y := (h^\Y |h^\Y \partial_i \Psi(f)),
\end{equation}
where the terms $(h^\Y|h^\Y)$, $(h^\Y|h^\Y \partial_i \ln \mathrm{A})$,
$(h^\Y|h^\Y \partial_i \Psi^\Y(f))$, $(h^\Y \partial_i \ln \mathrm{A}|h^\Y
\partial_j \ln \mathrm{A})$, and $(h^\Y \partial_i \Psi^\Y(f)|h^\Y \partial_j
\Psi^\Y(f))$ are given in Appendix~\ref{app:wavederivs}.  As referred to above,
the additional terms for inspiral signals associated with derivatives of the
intrinsic amplitude are contained in the $\mat{G}_{ij}$ and $\mat{J}_{i}$
terms.

Looking at the amplitude subspace maximization correction, $\R_{\alpha \mu i}
\M^{\alpha \beta} \R_{\beta \nu j}$, $\R_{\mu \nu i}$ has the block form
\begin{equation}\label{eq:Rblock}
\R_{\mu \nu i} = \left(\begin{array}{cc}
\hat{\R}_i & \tilde{\R}_i \\
-\tilde{\R}_i & \hat{\R}_i \end{array}\right)_{\mu \nu},
\end{equation}
where the blocks $\hat{\R}_i$ and $\tilde{\R}_i$ are defined as
\begin{equation}
\hat{\R}_i := \left(\begin{array}{cc}
R^{11}_i & R^{12}_i \\
R^{21}_i & R^{22}_i
\end{array}\right), \text{ and }
\tilde{\R}_i := \left(\begin{array}{cc}
R^{13}_i & R^{14}_i \\
R^{14}_i & R^{24}_i
\end{array}\right).
\end{equation}
These components are defined in Appendix~\ref{app:hterms}. As noted in Appendix
B of Ref.~\cite{PrixFstatMetric}, $\tilde{\R}_i$ contains only terms with
derivatives of the phase. However for the case of inspiral signals,
$\hat{\R}_i$ also contains terms with derivatives of both the antenna factors
and the intrinsic amplitude.  After using the symmetries of \eqref{eq:Minverse}
and \eqref{eq:Rblock}, and symmetrizing on $(i,j)$, the final form for
$\R_{\alpha \mu i} \M^{\alpha \beta} \R_{\beta \nu j}$ is
\begin{equation}\label{eq:Q}
\R_{\alpha \mu i} \M^{\alpha \beta} \R_{\beta \nu j} =
\left(\begin{array}{cccc}
Q^1_{ij} & Q^3_{ij} & 0 & Q^4_{ij} \\
Q^3_{ij} & Q^2_{ij} & -Q^4_{ij} & 0 \\
0 & -Q^4_{ij} & Q^1_{ij} & Q^3_{ij} \\
Q^4_{ij} & 0 & Q^3_{ij} & Q^2_{ij} \end{array}\right)_{\mu \nu}.
\end{equation}
Explicit expressions for the $Q$ components are given in
Appendix~\ref{app:hterms}.

Combining the terms from \eqref{eq:P} and \eqref{eq:Q}, we find the projected
Fisher matrix for inspiral signals has the same form as that of the
low-frequency limit of rotating neutron star signals (i.e., Appendix B of
Ref.~\cite{PrixFstatMetric}),
\begin{equation}\label{eq:mismatches}
\G_{\mu \nu i j} = \left(\begin{array}{cccc}
m^1_{ij} & m^3_{ij} & 0 & m^4_{ij} \\
m^3_{ij} & m^2_{ij} & -m^4_{ij} & 0 \\
0 & -m^4_{ij} & m^1_{ij} & m^3_{ij} \\
m^4_{ij} & 0 & m^3_{ij} & m^2_{ij} \end{array}\right)_{\mu \nu},
\end{equation}
where $m^k_{ij} = P^k_{ij} - Q^k_{ij}$. Combining \eqref{eq:fstatmetric} and
\eqref{eq:mismatches} gives the main result of this paper, namely, the coherent
\F-statistic metric for short-duration non-precessing inspiral signals.

It should be noted that, as in the rotating neutron star case, although the
\F-statistic metric \eqref{eq:fstatmetric} has projected out the amplitude
parameter subspace, it is still dependent on the amplitude parameters. This means
that what has been derived is actually a family of metrics that depend on the
the extrinsic parameters that enter the amplitude
parameters~\cite{PrixFstatMetric}. In order to produce a metric that is useful
for choosing template points to cover the parameter space, we must choose an
averaging procedure. As an example, Prix takes the average of the eigenvalues
of $(\M^{\alpha \beta} \G_{\alpha \beta i j} \Delta\lambda^i \Delta\lambda^j)$
to produce an average metric. This is motivated by the fact that this matrix
determines the extremal mismatches that can be obtained for any combination of
amplitude parameters~\cite{PrixFstatMetric, KrolakLRR2010}.

\section{Verification}
\label{sec:verif}

With the \F-statistic metric for short-duration non-precessing inspiral signals
in hand, we can verify its performance by comparing the fractional loss of the
\F-statistic for mismatched signals to that predicted by the metric. We do this
using a network of detectors corresponding to the locations and orientations of
the LIGO Hanford, LIGO Livingston, and Virgo detectors. The \acp{PSD} we use
for the LIGO detectors is the zero-detuning high-power advanced detector
configuration~\cite{advLIGO}. For Virgo, we use the advanced detector
\ac{PSD}~\cite{advVirgoPSD}. The waveform model used for this work is the
non-spinning restricted TaylorF2 \ac{PN} approximation~\cite{Damour:2001,
BIOPS}, which is given in Appendix~\ref{app:PNwaveform}. For computational
reasons, we start the waveforms at a low-frequency cutoff of 40Hz, although our
results should also be valid for other choices of the low-frequency cutoff.

\begin{figure*}
\centering
\subfloat[]{\includegraphics[]{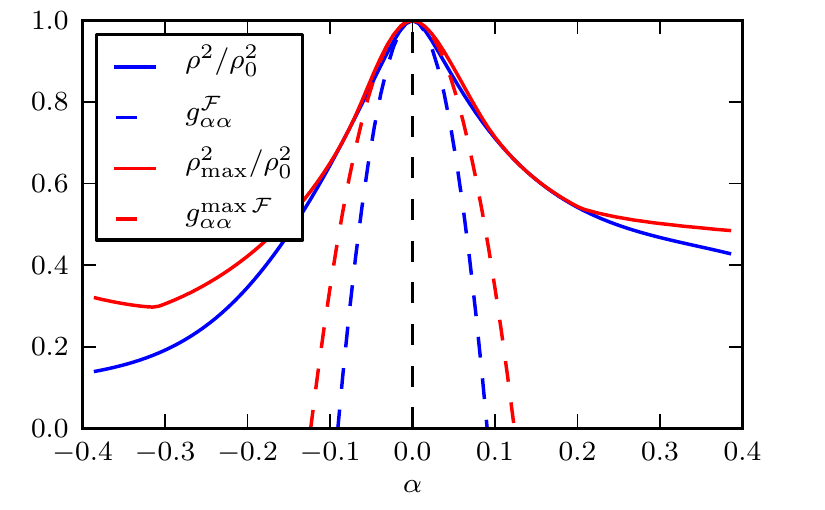}\label{fig:checkRA}}
\subfloat[]{\includegraphics[]{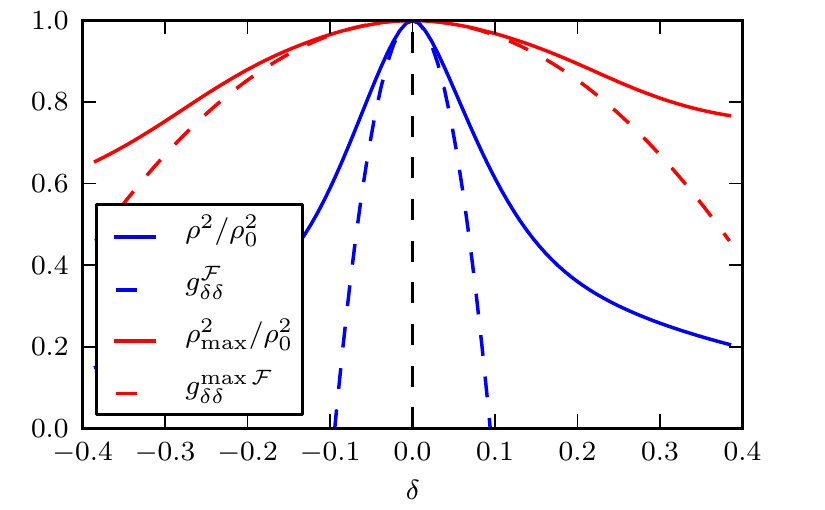}\label{fig:checkdec}}
\qquad
\subfloat[]{\includegraphics[]{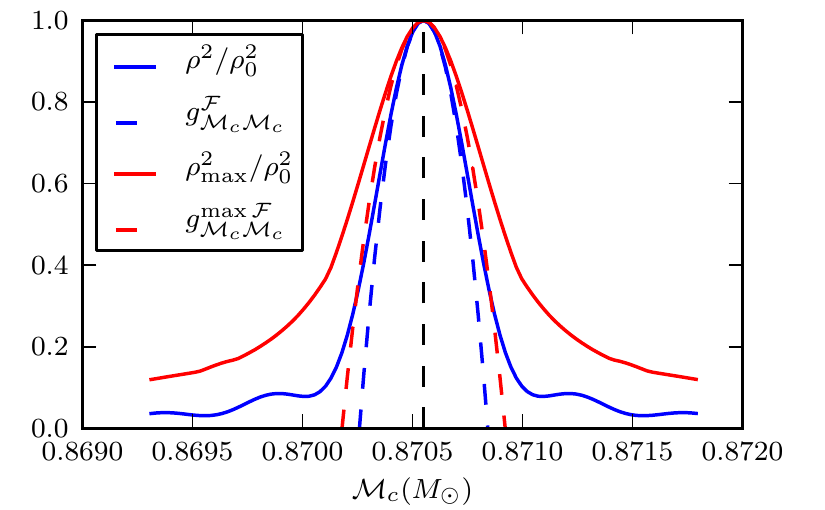}\label{fig:checkmchirp}}
\subfloat[]{\includegraphics[]{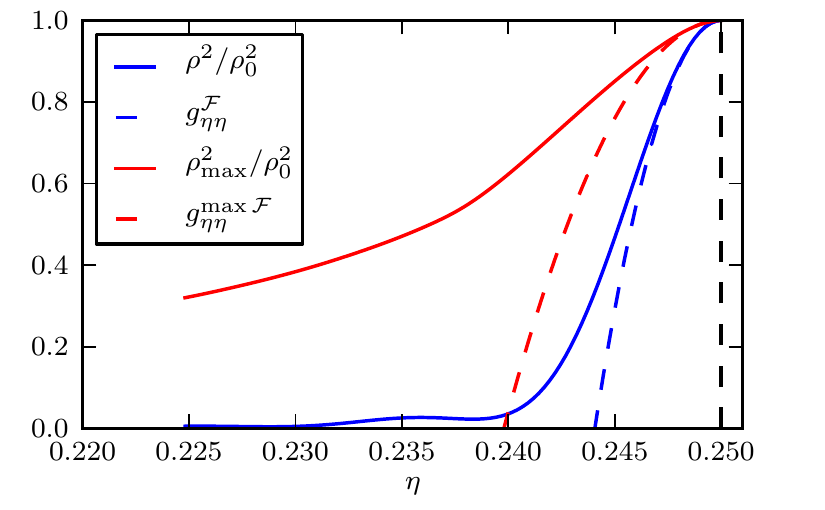}\label{fig:checketa}}
\qquad
\subfloat[]{\includegraphics[]{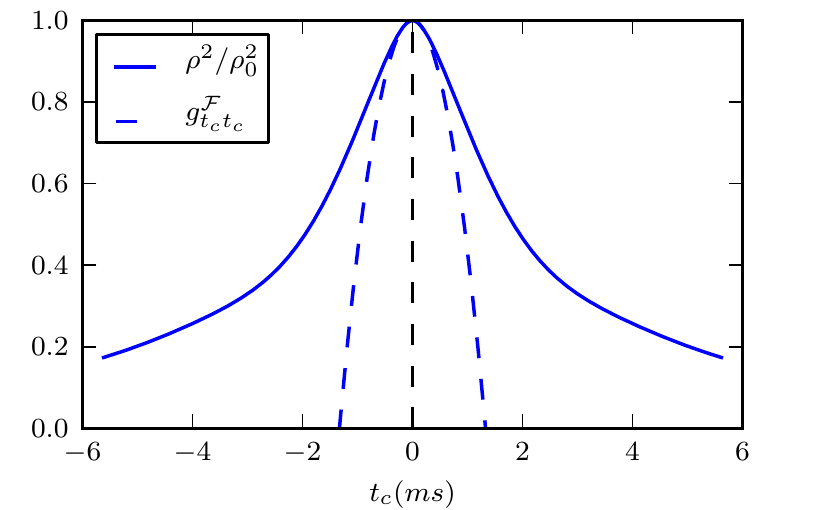}\label{fig:checkt}}
\caption{We show the fractional loss of \F{} as a function of parameter
mismatches for the coherent analysis, using the Advanced LIGO and Advanced
Virgo detectors, of an inspiral signal with component masses $m_1 = m_2 = 1.4
M_{\odot}$, sky location $(\alpha,\delta) = (0,0)$, and coalescence time $t_c =
0$. The vertical line shows the true parameters of the injected signal. The
solid lines show the observed fractional \F{} with and without maximization
over time (where appropriate). The dashed lines shown the predicted fractional
\F{} from the coherent metric, without ($g_{\mu \mu}^{\F}$) and with ($g_{\mu
\mu}^{\max \F}$) projection of the time dimension of the metric. The metric
accurately predicts the fractional loss of \F{} above a match of $\sim$0.95.
Panel~(a), (b), (c), (d), and (e) show the mismatch associated
variations of the right ascension, declination, chirp mass, symmetric mass
ratio, and coalescence time, respectively.}
\label{fig:checks}
\end{figure*}

We perform our tests using the following intrinsic parameters for the injected
signal: $m_1 = m_2 = 1.4 M_{\odot}$. The extrinsic parameters are:
$(\alpha,\delta) = (0,0)$, $t_c = 0$, $\phi_0=0$, $\psi = 0$, $\cos \iota = 1$,
and $\mathcal{D} = 200$ Mpc. The expected square coherent \ac{SNR} for this
signal is $2\F = 9.8^2$. We check the metric by computing the match, both with
and without maximization over time, while varying a single parameter. We do
this for the two intrinsic parameters $\{\M_c,\eta\}$, for the two extrinsic
sky-location parameters $\{\alpha,\delta\}$, and also for the time parameter
$\{t_c\}$. Figure~\ref{fig:checkRA} shows how the match varies when the
template's right ascension deviates from the signal's value, shown as the
vertical line. The metric reliably predicts the observed loss in \F{} above
$\sim$0.95.

\begin{figure*}
\centering
\subfloat[]{\includegraphics[]{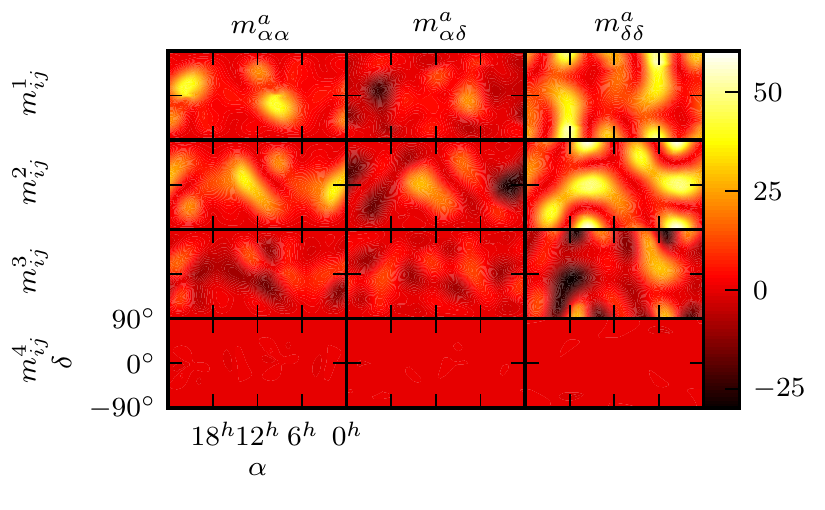}\label{fig:fullmismatches}}
\subfloat[]{\includegraphics[]{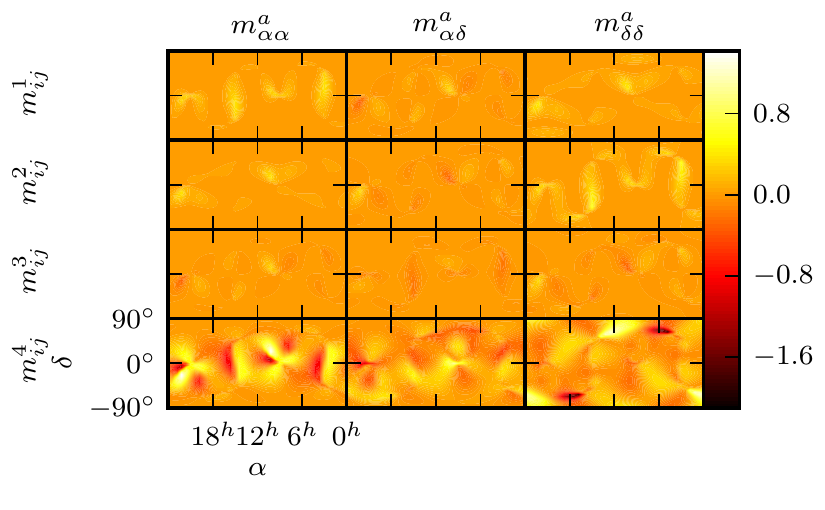}\label{fig:ampderivs}}
\caption{We show the separate mismatch components of the metric, $m^k_{ij}$ of
\eqref{eq:mismatches}, as functions of sky position in
(a).  The parts of $m^k_{ij}$ associated with the
derivatives of the detector response are shown in (b). For
the first three mismatches, these terms are typically more than an order of
magnitude smaller than the full mismatches.}
\label{fig:mismatches}
\end{figure*}

We are interested to see the effect that including derivatives of the detector
responses has on the metric calculation. To do this, first we check the
mismatches $m^a_{ij}$ from \eqref{eq:mismatches} associated with the
\F-statistic metric as a function of sky location, which can be seen in
Fig.~\ref{fig:fullmismatches}. Figure~\ref{fig:ampderivs} shows the portion of
these mismatches that originates from the derivatives of the detector
responses.  We see that for the first three mismatches, this portion is
typically an order of magnitude smaller than the full mismatch. As the fourth
mismatch is already an order of magnitude smaller than the first three,
including these terms is generally only a small correction to the metric.
However, as we shall see, there are points in parameter space where this is not
true.

\begin{figure}
\includegraphics[]{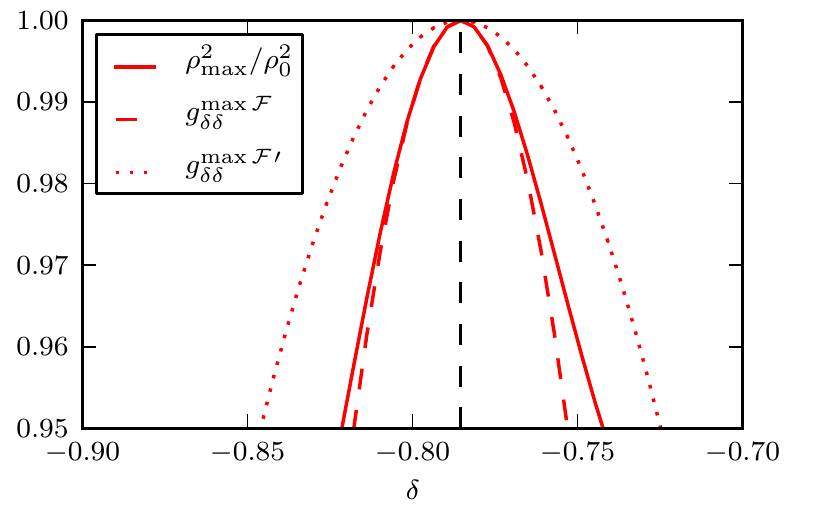}
\caption{We compare the time-maximized \F-statistic metric component with and
without the derivatives of the detector responses, denoted as $g_{\delta
\delta}^{\max \F}$ and $g_{\delta \delta}^{\max \F}{}'$, respectively, for a
specific set of parameter space coordinates: component masses $m_1 = m_2 = 1.4
M_{\odot}$, sky location $(\alpha,\delta) = (2.7, 0.5)$, and coalescence time
$t_c = 0$, reference phase $\phi_0=0$, polarization angle $\psi = 0$,
inclination angle $\cos \iota = 1$, distance $\mathcal{D} = 200$ Mpc. The
metric derived with the derivatives of detector responses better matches the
observed time-maximized fractional loss of \F{}.}
\label{fig:ampderivsdec}
\end{figure}

Finally, we check the effect of including the derivatives of the detector
responses in the metric in an extreme example. We use the following intrinsic
parameters for the injected signal: $m_1 = m_2 = 1.4 M_{\odot}$.  The extrinsic
parameters are: $(\alpha,\delta) = (0.785, -0.785)$, $t_c = 0$, $\phi_0=0$,
$\psi = 0$, $\cos \iota = 0$, $\mathcal{D} = 9.8$ Mpc. The distance is an order
of magnitude smaller than the previous comparisons in order to obtain an equal
expected square coherent \ac{SNR} for this signal, $2\F = 9.8^2$.
Figure~\ref{fig:ampderivsdec} compares the predictions from the metric derived
with and without the derivatives of detector responses to the observed
time-maximized fractional loss of \F{}. We see that the predictions from the
metric that includes the derivatives of the detector responses gives a
substantially better match to the observed time-maximized fractional loss of
\F{}.  However, it should be noted that the detector network is much less
sensitive to this point, which was chosen especially to show a large
discrepancy between including versus not including those derivatives. For the
majority of parameter space, the discrepancy is much smaller.

\section{Conclusion}

In this work, we derive the coherent \F-statistic metric associated with
short-duration non-precessing inspiral signals. This metric, understandably,
has very close ties to the coherent \F-statistic metric associated with
rotating neutron star signals.  However, in detail, there are several important
differences. For one, inspiral signals have a larger bandwidth, hence the
important single detector quantities are not the detectors' \ac{PSD} values at
a single frequency, but the integrated noise moments of the detectors'
\acp{PSD}. Secondly, the signal model includes intrinsic parameters in the
amplitude, which need to be properly accounted for in the metric derivation.

Even though this derivation closely follows that for the rotating neutron star
case, it includes previously ignored effects of the variation of the detector
responses. If desired, this could easily be incorporated into the rotating
neutron star coherent \F-statistic metric for a more complete picture of the
sky-tiling problem.

Important aspects that should be explored in the future include determining
other ways that the amplitude-dependent metric, derived here, can be
averaged~\cite{PrixFstatMetric} and applying the averaged metric to the
template covering problem associated with coherent searches short-duration
non-precessing inspiral signals. In order to efficiently perform this search,
it will need to be investigated how well the metric can be separated into an
intrinsic parameter space (e.g., the mass space) and an extrinsic parameter
space (e.g., the sky space) that could be tiled separately. This would allow
filters associated with different intrinsic parameters to be reused for the
extrinsic parameters that still need to be searched in a tiled
manner~\cite{Pai2000, Bose2011}.

Finally, because of the close ties between the metric and the projected Fisher
matrix, it may be interesting to use the derivation here to determine the sky
localization accuracy of a detector network, which could then be compared to
the derivations of~\cite{wen2010, Ajith:2009fz, Fairhurst2009, Fairhurst2010,
Fairhurst2012}.

\acknowledgments

The author would like to acknowledge many useful discussions with Sukanta Bose,
Badri Krishnan, and Reinhard Prix that this work is based upon. The author
would also like to thank Thomas Dent, Alex Nielsen, and Chris Pankow for useful
comments on this manuscript. The author is supported from the Max Planck
Gesellschaft.  Numerical overlap calculations in this work were accelerated
using pycuda~\cite{pycudapyopencl}.  This document has LIGO document number
LIGO-P1200091.

\appendix

\section{TaylorF2 PN Waveform}
\label{app:PNwaveform}

In this section we give the explicit formulae for the restricted \ac{SPA}
TaylorF2 inspiral waveform. As noted in Sec.~\ref{sec:approx}, the inspiral
waveform can be split into three pieces, a frequency-independent extrinsic
amplitude (i.e., a function of only extrinsic parameters), a
frequency-dependent intrinsic amplitude (i.e., a function of intrinsic
parameters and frequency), and a phase piece that depends on intrinsic
parameters, extrinsic parameters, and frequency.  The extrinsic amplitude for a
signal at distance $\mathcal{D}$ is given by,
\begin{equation}
h_0 = \sqrt{\frac{5}{24}} \frac{1}{\pi^{2/3}\mathcal{D}},
\end{equation}
and the intrinsic amplitude for a signal with chirp mass $\M_c$ is
\begin{equation}
\mathrm{A}(f) = \M_c^{-5/3} f^{-7/6}.
\end{equation}
For convenience, we define $\mathrm{A}$ without the frequency dependence as
\begin{equation}
\mathrm{A} := \M_c^{-5/3}.
\end{equation}
The phase of the inspiral waveform, expanded to 3.5~\ac{PN} order, can be
written as
\begin{multline}
\Psi^\Y(f) = 2 \pi f t^\Y - \frac{\pi}{4} + \phi_0 \\
+ \sum_{j=0}^{7} \psi_j f^{(-5+j)/3} \\
+ \sum_{j=5}^{6} \psi^l_j \ln(f) f^{(-5+j)/3},
\end{multline}
where $t^\Y := t_c - \vec{r}_\Y\cdot\hat{n}/c$ is a time parameter that
includes the time of arrival of the end of the waveform at the geocenter $t_c$
and the sky-location-dependent correction associated with a detector \Y's
location, and $\psi_j$ and $\psi^l_j$ are the phase coefficients associated
with the $j/2$~\ac{PN} order. These phase coefficients are given by
\begin{subequations} \begin{equation}
\psi_0 = \frac{3 \M_c^{-5/3}}{128 \pi^{5/3}},
\end{equation}
\begin{equation}
\psi_2 = \frac{5 \M_c^{-1} \eta^{-2/5}}{384 \pi} \left(\frac{743}{84} + 11
\eta\right),
\end{equation}
\begin{equation}
\psi_3 = \frac{-3 \pi^{1/3} \M_c^{-2/3} \eta^{-3/5}}{8},
\end{equation}
\begin{multline}
\psi_4 = \frac{5 \M_c^{-1/3} \eta^{-4/5}}{3072 \pi^{1/3}} \\
\times \left(\frac{3058673}{7056} + \frac{5429}{7} \eta + 617 \eta^2 \right),
\end{multline}
\begin{multline}
\psi_5 = \frac{5 \pi \eta^{-1}}{384} \left(\frac{7729}{84} - 13 \eta\right) \\
\times \left[1 + \log\left(6^{3/2} \pi \M_c \eta^{-3/5}\right)\right],
\end{multline}
\begin{equation}
\psi^l_5 = \frac{5 \pi \eta^{-1}}{384} \left(\frac{7729}{84} - 13 \eta\right),
\end{equation}
\begin{multline}
\psi_6 = \frac{\pi^{1/3} \M_c^{1/3} \eta^{-6/5}}{128}
\left(\frac{11583231236531}{1564738560} - 640 \pi^2 \right. \\
- \frac{6848}{7}\left[\gamma + \log\left(4 \pi^{1/3} \M_c^{1/3}
\eta^{-1/5}\right)\right] \\
+ \frac{5}{4}\left[\frac{-3147553127}{254016} + 451 \pi^2\right] \eta \\
\left.+ \frac{76055}{576} \eta^2 - \frac{127825}{432} \eta^3\right),
\end{multline}
\begin{equation}
\psi^l_6 = \frac{-107 \pi^{1/3} \M_c^{1/3} \eta^{-6/5}}{42},
\end{equation}
\begin{multline}
\psi_7 = \frac{5 \pi^{5/3} \M_c^{2/3} \eta^{-7/5}}{32256} \\
\times \left(\frac{15419335}{336} + \frac{75703}{2} \eta - 14809 \eta^2\right).
\end{multline}\end{subequations}
Any \ac{PN} coefficients of 3.5~\ac{PN} order or lower not defined above are
identically zero.

\section{Explicit expressions for detector polarization response derivatives}
\label{app:detpolrespderivs}

Based on the expressions for the detector polarization responses in
Sec.~\ref{sec:signal} and \cite{PrixFstatTechNote}, the derivatives of the detector
polarization responses can be obtained in terms of derivatives of the
polarization-independent basis tensors,
\begin{equation}
\partial_a F_{+,\cross}^Y = \partial_a \epsilon_{+,\cross}^{ij} d_{ij}^\Y.
\end{equation}
These in turn can be written in terms of derivatives of the radiation frame
basis vector,
\begin{subequations}\begin{multline}
\partial_a \epsilon_{+}^{ij} = \left\{ (\partial_a \vec{\xi}) \otimes \vec{\xi}
+ \vec{\xi} \otimes (\partial_a \vec{\xi}) \right.\\
\left.- (\partial_a \vec{\eta}) \otimes \vec{\eta} - \vec{\eta} \otimes
(\partial_a \vec{\eta}) \right\}^{ij},
\end{multline}
\begin{multline}
\partial_a \epsilon_{\cross}^{ij} = \left\{ (\partial_a \vec{\xi}) \otimes
\vec{\eta} + \vec{\xi} \otimes (\partial_a \vec{\eta}) \right.\\
\left.+ (\partial_a \vec{\eta}) \otimes \vec{\xi} + \vec{\eta} \otimes
(\partial_a \vec{\xi}) \right\}^{ij}.
\end{multline}\end{subequations}
The explicit formulae for the derivatives of the radiation frame basis vectors
with respect to the right ascension and declination are given as
\begin{subequations}\begin{equation}\label{eq:dndRA}
\partial_\RA \hat{n} = (-\cos{\dec}\sin{\RA}, \cos{\dec}\cos{\RA}, 0),
\end{equation}
\begin{equation}\label{eq:dnddec}
\partial_\dec \hat{n} = (-\sin{\dec}\cos{\RA}, -\sin{\dec}\sin{\RA},
\cos{\dec}),
\end{equation}
\begin{equation}
\partial_\RA \hat{\xi} = (\cos{\RA}, \sin{\RA}, 0),
\end{equation}
\begin{equation}
\partial_\dec \hat{\xi} = (0, 0, 0),
\end{equation}
\begin{equation}
\partial_\RA \hat{\eta} = (\sin{\dec}\sin{\RA}, -\sin{\dec}\cos{\RA}, 0),
\end{equation}
\begin{equation}
\partial_\dec \hat{\eta} = (-\cos{\dec}\cos{\RA}, -\cos{\dec}\sin{\RA},
-\sin{\dec}).
\end{equation}\end{subequations}

\section{Explicit expressions for inner product derivatives}
\label{app:wavederivs}

In this section we define the inner products that are needed for the coherent
\F-statistic for short-duration non-precessing inspiral signals in terms of
derivatives of intrinsic and extrinsic parameters and combinations of detector
noise moment integrals, given by~\eqref{eq:noisemoment} of
Appendix~\ref{app:moments}.

The simplest inner product required, which contains no derivatives, is used by
$\M_{\mu \nu}$ and $\mat{H}$,
\begin{equation}
(h^\Y|h^\Y) = \mathrm{A}^2 I(7,0)^\Y =: H^\Y.
\end{equation}
The inner product that contains a single derivative of the intrinsic amplitude
and is used by $\mat{J}_i$ is
\begin{equation}
(h^\Y|h^\Y \partial_i \ln \mathrm{A}) = -\frac{5}{3} \frac{\partial \ln
\M_c}{\partial \lambda^i} \mathrm{A}^2 I(7,0)^\Y =: J^\Y_i.
\end{equation}
The inner product that contains a single derivative of the phase and is used by
$\mat{K}_i$ is
\begin{multline}
(h^\Y|h^\Y \partial_i \Psi^\Y(f)) = \mathrm{A}^2 \Biggl(\frac{\partial
t^\Y}{\partial \lambda^i} 2\pi I(4,0)^\Y \\
+ \sum_k \left[\frac{\partial \psi_k}{\partial \lambda^i} I(12-k,0)^\Y +
\frac{\partial \psi^l_k}{\partial \lambda^i} I(12-k,1)^\Y \right]\Biggr) =:
K^\Y_i.
\end{multline}
Finally, the inner product that contains both two single derivatives of the
intrinsic amplitude and two single derivatives of the phase is used by
$\mat{G}_{ij}$,
\begin{equation}
(h^\Y \partial_i \ln \mathrm{A}|h^\Y \partial_j \ln \mathrm{A}) + (h^\Y
\partial_i \Psi^\Y(f)|h^\Y \partial_j \Psi^\Y(f)) =: \mat{G}_{ij}.
\end{equation}
These individual inner products are given by
\begin{multline}
(h^\Y \partial_i \ln \mathrm{A}|h^\Y \partial_j \ln \mathrm{A}) = \\
\frac{25}{9} \frac{\partial \ln \M_c}{\partial \lambda^i} \frac{\partial \ln
\M_c}{\partial \lambda^j} \mathrm{A}^2 I(7,0)^\Y,
\end{multline}
and
\begin{multline}
(h^\Y \partial_i \Psi^\Y(f)|h^\Y \partial_j \Psi^\Y(f)) = \mathrm{A}^2
\biggl[\frac{\partial t^\Y}{\partial \lambda^i} \frac{\partial t^\Y}{\partial
\lambda^j} 4\pi^2 I(1,0)^\Y \\
+ \sum_{k} \left( \frac{\partial t^\Y}{\partial \lambda^i} \frac{\partial
\psi_k}{\partial \lambda^j}
+ \frac{\partial \psi_k}{\partial \lambda^i} \frac{\partial t^\Y}{\partial
\lambda^j}\right) 2 \pi I(9-k,0)^\Y \\
+ \sum_{k} \left( \frac{\partial t^\Y}{\partial \lambda^i} \frac{\partial
\psi^l_k}{\partial \lambda^j}
+ \frac{\partial \psi^l_k}{\partial \lambda^i} \frac{\partial t^\Y}{\partial
\lambda^j} \right) 2 \pi I(9-k,1)^\Y \\
+ \sum_{k,l} \frac{\partial \psi_k}{\partial \lambda^i} \frac{\partial
\psi_l}{\partial \lambda^j} I(17-k-l,0)^\Y \\
+ \sum_{k,l} \frac{\partial \psi_k}{\partial \lambda^i} \frac{\partial
\psi^l_l}{\partial \lambda^j} I(17-k-l,1)^\Y \\
+ \sum_{k,l} \frac{\partial \psi^l_k}{\partial \lambda^i} \frac{\partial
\psi^l_l}{\partial \lambda^j} I(17-k-l,2)^\Y\biggr].
\end{multline}

\section{Explicit terms associated with $\G_{\mu \nu i j}$}
\label{app:hterms}

The explicit formulae the amplitude subspace maximization correction $Q$ of the
projected Fisher matrix, written in terms of the $R^{\mu \nu}_i$ components,
are given as
\begin{subequations}\begin{multline}
D Q^1_{ij} = A (R^{21}_i R^{21}_j + R^{14}_i R^{14}_j) + B (R^{11}_i R^{11}_j +
R^{13}_i R^{13}_j) \\
- C (R^{11}_i R^{21}_j + R^{21}_i R^{11}_j + R^{13}_i R^{14}_j + R^{14}_i
R^{13}_j),
\end{multline}\begin{multline}
D Q^2_{ij} = A (R^{22}_i R^{22}_j + R^{24}_i R^{24}_j) + B (R^{12}_i R^{12}_j +
R^{14}_i R^{14}_j) \\
- C (R^{12}_i R^{22}_j + R^{22}_i R^{12}_j + R^{14}_i R^{24}_j + R^{24}_i
R^{14}_j),
\end{multline}\begin{multline}
D Q^3_{ij} = A (R^{21}_i R^{22}_j + R^{22}_i R^{21}_j + R^{14}_i R^{24}_j +
R^{24}_i R^{14}_j) \\
+ B (R^{11}_i R^{12}_j + R^{12}_i R^{11}_j + R^{13}_i R^{14}_j + R^{14}_i
R^{13}_j) \\
- C (R^{11}_i R^{22}_j + R^{22}_i R^{11}_j + R^{12}_i R^{21}_j + R^{21}_i
R^{12}_j \\
+ R^{13}_i R^{24}_j + R^{24}_i R^{13}_j + 2 R^{14}_i R^{14}_j),
\end{multline}\begin{multline}
D Q^4_{ij} = A (R^{21}_i R^{24}_j + R^{24}_i R^{21}_j - R^{14}_i R^{22}_j -
R^{22}_i R^{14}_j) \\
+ B (R^{11}_i R^{14}_j + R^{14}_i R^{11}_j - R^{13}_i R^{12}_j - R^{12}_i
R^{13}_j) \\
- C (R^{11}_i R^{24}_j + R^{24}_i R^{11}_j + R^{14}_i R^{21}_j + R^{21}_i
R^{14}_j \\
- R^{13}_i R^{22}_j - R^{22}_i R^{13}_j - R^{12}_i R^{14}_j - R^{14}_i
R^{12}_j),
\end{multline}\end{subequations}
where we recall that $A,B,C,D$ come from \eqref{eq:Minverse} for $\M^{\mu\nu}$.
These $R^{ij}_k$ components are given by
\begin{subequations}\begin{equation}
R^{11}_i = \mat{f}^{++}_i \cdot \mat{H} + \mat{f}^{++} \cdot \mat{J}_i,
\end{equation}\begin{equation}
R^{12}_i = \mat{f}^{\cross+}_i \cdot \mat{H} + \mat{f}^{\cross+} \cdot
\mat{J}_i,
\end{equation}\begin{equation}
R^{21}_i = \mat{f}^{+\cross}_i \cdot \mat{H} + \mat{f}^{+\cross} \cdot
\mat{J}_i,
\end{equation}\begin{equation}
R^{22}_i = \mat{f}^{\cross\cross}_i \cdot \mat{H} + \mat{f}^{\cross\cross}
\cdot \mat{J}_i,
\end{equation}\begin{equation}
R^{13}_i = \mat{f}^{++} \cdot \mat{K}_i,
\end{equation}\begin{equation}
R^{14}_i = \mat{f}^{+\cross} \cdot \mat{K}_i,
\end{equation}\begin{equation}
R^{24}_i = \mat{f}^{\cross\cross} \cdot \mat{K}_i.
\end{equation}\label{eq:Rs}\end{subequations}

It is interesting to note that all of the $Q$ components contain terms
associated with derivatives of the detector polarization responses as well as
terms associated with derivatives of the intrinsic amplitude.

\section{Detector PSD Moment Integrals}
\label{app:moments}

In this section define the noise moment integrals $I(k,l)^\Y$ of detector \Y's
\ac{PSD},
\begin{multline}\label{eq:noisemoment}
I(k,l)^\Y := \frac{1}{\mathrm{A}^2}(h^\Y| h^\Y \ln^l(f) f^{-k/3}) \\
= \int\limits_{f_{\rm low}}^{f_{\rm high}} \frac{\ln^l(f)
f^{-k/3}}{S^{\Y}(f)} df.
\end{multline}
This is the same definition as in~\cite{Keppel2012}. Based on the definition in
\eqref{eq:noisemoment}, it is easy to see that powers of the frequency in the
inner-product can be manipulated as $(h^\Y  \ln^l_1(f) f^{-k_1/3}| h^\Y
\ln^l_2(f) f^{-k_2/3}) = (h^\Y| h^\Y \ln^{l_1+l_2}(f) f^{-(k_1+k_2)/3}) =
\mathrm{A}^2 I(k_1+k_2+7,l_1+l_2)^\Y$.

The detector \ac{PSD} moments required for the metric calculation associated
with restricted \ac{SPA} TaylorF2 inspiral waveforms expanded to 3.5~\ac{PN}
order are
\begin{align}
(k,l) \in \{&(1,0), (2,0), (3,0), (4,0), (5,0), (6,0), \nonumber\\
&(7,0), (8,0), (9,0), (10,0), (11,0), \nonumber\\
&(12,0), (13,0), (14,0), (15,0), (17,0), \nonumber\\
&(3,1), (4,1), (5,1), (6,1), (7,1), \nonumber\\
&(8,1), (9,1), (10,1), (11,1), (12,1), \nonumber\\
&(5,2), (6,2), (7,2)\}. \nonumber
\end{align}

\section{Derivatives of \ac{PN} Coefficients}
\label{ap:PNcoeffderivs}

Here we give explicit expressions for the derivatives of the \ac{PN}
coefficients associated with the phase in terms of the symmetric mass ratio
$\eta$ and the chirp mass $\mathcal{M}$, as computed in~\cite{Keppel2012}.
First, the derivatives with respect to $\mathcal{M}$,
\begin{equation}
\partial_{\mathcal{M}} \psi_{0} = \frac{-5}{128 \pi^{5/3} \mathcal{M}^{8/3}},
\end{equation}
\begin{equation}
\partial_{\mathcal{M}} \psi_{2} = \frac{-5}{384 \pi \mathcal{M}^2 \eta^{2/5}}
\left(\frac{743}{84} + 11 \eta\right),
\end{equation}
\begin{equation}
\partial_{\mathcal{M}} \psi_{3} = \frac{\pi^{1/3}}{4 \mathcal{M}^{5/3}
\eta^{3/5}},
\end{equation}
\begin{multline}
\partial_{\mathcal{M}} \psi_{4} = \frac{-5}{9216 \pi^{1/3} \mathcal{M}^{4/3}
\eta^{4/5}} \\ \times \left(\frac{3058673}{7056} + \frac{5429}{7} \eta + 617
\eta^2 \right),
\end{multline}
\begin{equation}
\partial_{\mathcal{M}} \psi_{5} = \frac{5 \pi}{384 \mathcal{M} \eta}
\left(\frac{7729}{84} - 13 \eta\right),
\end{equation}
\begin{multline}
\partial_{\mathcal{M}} \psi_{6} = \frac{\pi^{1/3}}{384 \mathcal{M}^{2/3}
\eta^{6/5}} \left(\frac{10052469856691}{1564738560} \right.  \\
\left.- 640 \pi^2 - \frac{6848}{7}\left[\gamma + \ln\left(4 \pi^{1/3}
\mathcal{M}^{1/3} \eta^{-1/5}\right)\right] \right. \\
\left.+ \frac{5}{4}\left[\frac{-3147553127}{254016} + 451 \pi^2\right] \eta
\right. \\
\left.+ \frac{76055}{576} \eta^2 - \frac{127825}{432} \eta^3\right),
\end{multline}
\begin{equation}
\partial_{\mathcal{M}} \psi^l_{6} = \frac{-107 \pi^{1/3}}{126 \mathcal{M}^{2/3}
\eta^{6/5}},
\end{equation}
\begin{multline}
\partial_{\mathcal{M}} \psi_{7} = \frac{5 \pi^{5/3}}{48384 \mathcal{M}^{1/3}
\eta^{7/5}} \\
\times \left(\frac{15419335}{336} + \frac{75703}{2} \eta - 14809 \eta^2\right).
\end{multline}
Now the derivatives with respect to $\eta$,
\begin{equation}
\partial_{\eta} \psi_{2} = \frac{-1}{384 \pi \mathcal{M} \eta^{7/5}}
\left(\frac{743}{42} - 33 \eta\right),
\end{equation}
\begin{equation}
\partial_{\eta} \psi_{3} = \frac{9 \pi^{1/3}}{40 \mathcal{M}^{2/3} \eta^{8/5}},
\end{equation}
\begin{multline}
\partial_{\eta} \psi_{4} = \frac{-3}{3072 \pi^{1/3} \mathcal{M}^{1/3}
\eta^{9/5}} \\ \times \left(\frac{3058673}{5292} - \frac{5429}{21} \eta + 1234
\eta^2 \right),
\end{multline}

\begin{multline}
\partial_{\eta} \psi_{5} = \frac{-\pi}{384 \eta^2} \\ \times
\left(\frac{7729}{84} \left[8 + 5 \ln\left(6^{3/2} \pi \mathcal{M}
\eta^{-3/5}\right)\right] - 39 \eta\right),
\end{multline}
\begin{equation}
\partial_{\eta} \psi^l_{5} = \frac{-38645 \pi}{32256 \eta^2},
\end{equation}
\begin{multline}
\partial_{\eta} \psi_{6} = \frac{-\pi^{1/3} \mathcal{M}^{1/3}}{640 \eta^{11/5}}
\left(\frac{11328104339891}{260789760} - 3840 \pi^2 \right. \\
\left.- \frac{41088}{7}\left[\gamma + \ln\left(4 \pi^{1/3} \mathcal{M}^{1/3}
\eta^{-1/5}\right)\right] \right. \\
\left.+ \frac{5}{4}\left[\frac{-3147553127}{254016} + 451 \pi^2\right] \eta
\right. \\
\left.- \frac{76055}{144} \eta^2 + \frac{127825}{48} \eta^3\right),
\end{multline}
\begin{equation}
\partial_{\eta} \psi^l_{6} = \frac{107 \pi^{1/3} \mathcal{M}^{1/3}}{35
\eta^{11/5}},
\end{equation}
\begin{multline}
\partial_{\eta} \psi_{7} = \frac{-\pi^{5/3} \mathcal{M}^{2/3}}{32256
\eta^{12/5}} \\
\times \left(\frac{15419335}{48} + 75703\eta + 44427 \eta^2\right).
\end{multline}
Any derivatives of the coefficients of 3.5~\ac{PN} order or lower not defined above
are identically zero.

The phase also contains a term associated with the relative time shift between
the arrival of a signal at a detector's location compared to the arrival of the
signal at the geocenter. These derivatives are as follows,
\begin{subequations}\begin{equation}
\partial_\RA t^Y = - \frac{\vec{r}_\Y\cdot\partial_\RA\hat{n}}{c},
\end{equation}\begin{equation}
\partial_\dec t^Y = - \frac{\vec{r}_\Y\cdot\partial_\dec\hat{n}}{c},
\end{equation}\end{subequations}
where $\partial_\RA\hat{n}$ and $\partial_\dec\hat{n}$ are given by
\eqref{eq:dndRA} and \eqref{eq:dnddec}, respectively.

\bibliography{references}
\end{document}